\pdfoutput=1

\documentclass[10pt]{article}

\usepackage{authblk}
\usepackage[]{naacl2021}

\usepackage{times}
\usepackage{latexsym}

\usepackage[T1]{fontenc}

\usepackage{microtype}

\usepackage{amsmath}
\usepackage{amsfonts} 
\usepackage{multirow}
\usepackage{amssymb}
\usepackage{booktabs}

\usepackage{caption}
\usepackage{subcaption}
\usepackage{graphicx}
\usepackage{comment}
\usepackage{array}
\usepackage{lmodern}
\usepackage{flushend}

\newcolumntype{P}[1]{>{\raggedright\arraybackslash}p{#1}}

\newcommand{\ignore}[1]{}

\title{Case Study on Detecting COVID-19 Health-Related Misinformation in Social Media}

\author{
\begin{tabular}{c}
Mir Mehedi Ahsan Pritom$^{1}$, Rosana Montanez Rodriguez$^{1}$, 
Asad Ali Khan$^{2}$, Sebastian A. Nugroho$^{2}$, \\
Esra'a Alrashydah$^{3}$, Beatrice N. Ruiz$^{4}$, Anthony Rios$^{5}$
\end{tabular}%
\\
\begin{tabular}{cc}
    \textmd{Department of Computer Science$^{1}$} &  \textmd{Department of Civil and Environmental Engineering$^{3}$} \\
    \textmd{Department of Electrical and Computer Engineering$^{2}$} & \textmd{Department of Psychology$^{4}$}
\end{tabular}

\\
\textmd{Department of Information Systems and Cyber Security$^{5}$}
\\
\textnormal{University of Texas at San Antonio, USA}
\\
\textnormal{Email: \{mirmehedi.pritom, rosana.montanezrodriguez, asad.khan, sebastian.nugroho, anthony.rios\}@utsa.edu}
}

\ignore{
\author[1]{{Mir Mehedi Ahsan Pritom}} \author[1]{Rosana Montanez Rodriguez} 
\author[2]{Asad Ali Khan}
\author[2]{{Sebastian  A. Nugroho}\\}
\author[3]{Esra'a Alrashydah}
\affil[1]{Department of Computer Science, University of Texas at San Antonio}
\affil[2]{Department of Electrical and Computer Engineering, University of Texas at San Antonio}
\affil[3]{Department of Civil and Environmental Engineering, University of Texas at San Antonio}
\affil[4]{Department of Psychology, University of Texas at San Antonio}
\affil[5]{Department of Information Systems \& Cyber Security, University of Texas at San Antonio}
}

\vspace{4cm}

\begin{document}

\maketitle

\begin{abstract}
\ignore{
Countering COVID-19 health-related misinformation poses a unique challenge that typical health campaigns do not encounter, which is {\color{green}trying to remove the continued influence of misinformation}\footnote{Are we addressing the psychological research part? If not, then why mentioning about continued influence part?}. Unlike previous research on COVID-19 misinformation, 

COVID-19 pandemic is causing unprecedented physical and mental health problems. Within the timeline of this pandemic there has been an infodemic of misinformation going on as well \cite{covid_infodemic}. In this era of social media, this infodemic has introduced us to a new challenge of tackling health-related misinformation campaigns over social media. With the increasing volume of social media users during this pandemic \cite{social_media_user_increase}, and large number of health-related misinformation flying through social media platforms, it is worth exploring the problem of COVID-19 health-related misinformation dissemination in detail to discover ways to tackle this issue more proactively in coming days. }

COVID-19 pandemic has generated what public health officials called an infodemic of misinformation. As social distancing and stay-at-home orders came into effect, many turned to social media for socializing. This increase in social media usage has made it a prime vehicle for the spreading of misinformation. This paper presents a mechanism to detect COVID-19 health-related misinformation in social media following an interdisciplinary approach. Leveraging social psychology as a foundation and existing misinformation frameworks, we defined misinformation themes and associated keywords incorporated into the misinformation detection mechanism using applied machine learning techniques. Next, using the Twitter dataset, we explored the performance of the proposed methodology using multiple state-of-the-art machine learning classifiers. Our method shows promising results with at most 78\% accuracy in classifying health-related misinformation versus true information using uni-gram-based NLP feature generations from tweets and the Decision Tree classifier. We also provide suggestions on alternatives for countering misinformation and ethical consideration for the study.

\ignore{In this paper, we address the existing multidisciplinary techniques that help us understand and detect COVID-19 themed health-related misinformation.those are ongoing in popular online social media throughout this pandemic. Our objective is to contribute to advancing methods to counter COVID-19 health-related misinformation spread in social media by presenting a detection method based on the proposed misinformation theory

In this pilot study, we chose Twitter as our source social media platform to collect COVID-19 health-related tweets. We propose new research methodologies to address this urgent problem by listing the relevant ``themes and keywords'' for filtering only health-related tweets. To the best of our knowledge there is no ground-truth labeled dataset available for this sort of analysis, which is also a key drawback in the current settings. In this study, we manually labeled tweets for health-related misinformation versus true information, which is used to train the supervised learning classifiers. }  

\ignore{and {\color{red}(x)\%}  \footnote{\color{red} From Sebastian to Rosa: I could not find the FP rate in the spreadsheet.} false positive rates with simple ML models. } 

\end{abstract}

\section{Introduction}

The ongoing COVID-19 pandemic has brought an unprecedented health crisis.\ignore{ The pandemic, which emerged for the first time as early as December 2019 in Wuhan, China~\cite{Wu2020}, is caused by the virus SARS-CoV-2, which attacks human respiratory systems~\cite{Brosnahan}. As of March 2021 in the US, there are 28,355,420 active cases and 510,777 deaths ~\cite{CDC_US_toll}.} Along with the physical health-related effects, the pandemic has brought changes to daily social interactions\ignore{that had to affect social interactions, that have increase stress and mental health issues,} 
such as teleworking, social distancing, and stay-at-home orders leading to high usage of social media \cite{social_media_user_increase}.\ignore{ which is exacerbated by the death or illness of loved ones.} These conditions have paved the way for opportunistic bad guys to {\em fish in the troubled waters}. From the beginning of the COVID-19 pandemic, an excessive amount of misinformation has spread across social and online digital media~\cite{BARUA2020100119, socialmedia_infodemic}. These misinformation campaigns include hoaxes, rumors, propaganda, or conspiracy theories, often with themed products or services that may protect from contracting the COVID-19 infection ~\cite{Mir_COVID_MisinfoAttack, who_infodemic}. Social media became the main conduit for the spread of COVID-19 misinformation. 
The abundance of health-related misinformation spreading over social media presents a threat to public health, especially in controlling and mitigating the spread of COVID-19~\cite{Islam2020astmh}. Misinformation campaigns also affect public attitudes towards health guidance compliance and hampering efforts towards preventing the spread. In some cases, individuals have lost their lives by making decisions based on misinformation~\cite{BARUA2020100119}. Therefore, it is imperative to combat COVID-19 health-related misinformation to minimize its adverse impact on public health~\cite{Ali2020}.

In the past, researchers have focused on identifying social media misinformation \cite{ijcai2017_misinfo} using various machine learning (ML) and deep learning-based approaches.
\ignore{{\color{red}Despite these developments, ML algorithms' effectiveness in detecting health-related COVID-19 misinformation is unfortunately not very well understood. Moreover, there is a lack of ground-truth and validated datasets to verify model accuracy for these research efforts.  }} 
However, the effectiveness of those approaches in tackling health-related COVID-19 misinformation from social media is unknown. 
There is also a lack of ground-truth and validated datasets to verify model accuracy for the previous research efforts. Moreover, we find lacking in the existing literature to address the misinformation problem with coordinated interdisciplinary approaches from social psychology, information operations, and data science (i.e., applied machine learning). Motivated by the lacking, the primary objective of this study is to leverage interdisciplinary techniques to understand the COVID-19 health-related misinformation problem and derive insights by detecting misinformation using the Natural Language Processing (NLP) methods and Machine Learning (ML) classifiers. 
Grounded in existing work on misinformation propagation \cite{schneider2019miss} and source credibility theory \cite{hovland1951influence} in social psychology, our study 
derives various credible health-related {\em themes} 
for understanding health misinformation and propose detection utilizing state-of-the-art techniques from NLP and applied machine learning.

\thispagestyle{plain}

We use {\em Twitter} as our social media platform to study the effectiveness of our proposed methodology.
We have collected, processed, labeled, and analyzed tweets to train and test the supervised ML classifiers. Finally, we have analyzed the performance of the classifiers following our mechanism using standard performance metrics such as {\em accuracy}, {\em precision}, {\em recall}, {\em F1-score}, and {\em Macro-F1-score}. The major \textbf{contributions of this paper} are, 
\begin{itemize}
    \item Provide a detailed methodology with a prototype for detecting COVID-19 health-related misinformation from social media (i.e., Twitter).
    \item Propose a more fair social media (e.g., Twitter) annotation process for labeling misinformation. 
    \item Provide a labeled ground-truth dataset for COVID-19 health-related tweets for future model verification.
    \item Provide the efficacy of state-of-the-art classifiers to detect COVID-19 health-related misinformation tweets leveraging different NLP text representation methods such as Bag-of-Words (BoW) and $n$-gram.
\end{itemize}

\thispagestyle{plain}






\noindent{\bf Paper outline}. 
Section \ref{sec:related_work} presents the problem background motivation and related works. 
Section \ref{sec:methodology} presents the research methodology, experiment details with the dataset collection, processing, and analyzing steps.
Section \ref{sec:discuss}  discusses the experiment results, limitations, future research directions, and  
ethical considerations of the present study. Section \ref{sec:conclusion} concludes the paper.

\section{Background and Related Works}
\label{sec:related_work}


\ignore{
The literature reported: 
[1]	Different platforms to collect data. (in our proposed study we will use Twitter)
[2]	Different techniques to detect misinformation (in our proposal, we will be using machine learning)
[3]	Different features to help detecting misinformation 
[4]	Different information theories to measure information uncertainty ( I agree with Mir suggestions, Entropy and mutual information )}




We find that misinformation research has also been driven by the COVID-19 pandemic as there are lots of ongoing research on fake news and social media misinformation. In this section, we draw from the previous works on how existing misinformation detection research is leveraging Natural Language Processing (NLP), Machine Learning (ML), and interdisciplinary techniques such as information kill chain (e.g., step for the propagation of misinformation), and social psychology. 

\ignore{Different techniques were reported and used to analyze and detect COVID-19 health-related misinformation. Some of these studies used feature engineering and introduced new features that could help to detect misinformation~\cite{Serrano2020}. Other research studies were conducted to detect uncertain information using information concept theories, such as entropy and mutual information~\cite{Husari2018}.}

\paragraph{\textbf{Background Motivation.}}
Authors in ~\cite{pan2020fighting, sylvia2020prologue} 
highlights how the COVID-19 infodemic has added additional challenges for the public health community. Infodemic is the product of an overabundance of information that undermines public health efforts to address the pandemic \cite{who_infodemic}. The effect of COVID-19 misinformation also impacts law enforcement and public safety entities \cite{gradon2020crime}. The study finds that social media have a higher prevalence of misinformation than news outlets \cite{bridgman2020causes}. Another study highlights that an increase in Twitter conversation on COVID-19 is also a good predictor of COVID-19 regional contagion \cite{singh2020first}. Authors in \cite{roozenbeek2020susceptibility} and \cite{bridgman2020causes} report that exposure to misinformation increases individual's misconception on COVID-19 and lowers their compliance with public health prevention guidelines. 
 
Our approach is also influenced by Information Operations Kill Chain ~\cite{schneider2019miss}. The framework is based on the Russian ``Operation Infektion” misinformation campaign and provides the basis for our focus on existing grievances. A critical characteristic of misinformation is that it propagates using existing channels by aligning the messages to pre-existing grievances and beliefs in a group~\cite{schneider2019miss, fbi2019mis}. Using existing media associated with a credible source (i.e., credible from the audience perspective) makes the message more likely to be accepted by the audience \cite{hovland1951influence}. Moreover, \citet{Islam2020astmh} reveals that the oversupply of health-related misinformation fueled by rumors, stigma, and conspiracy theories in social media platforms provides critical, adverse implications towards individuals and communities.

\paragraph{\textbf{COVID-19 Misinformation.}}
Studies comparing the performance of different ML algorithms have been conducted in the literature. 
For instance, \cite{CHOUDRIE2021106716}  analyzes how older adults process various kinds of infodemic about COVID-19 prevention and cure using Decision Tree and Convolutional Neural Network techniques. Although this study focuses on COVID-19 health-related misinformation, the data is collected via online interviews with 20 adults.  \cite{Mackey2021} have presented an application of unsupervised learning to detect misinformation on Twitter using ``hydroxychloroquine'' as the keyword. However, the study has only scoped to detect misinformation related to the word ``hydroxychloroquine," one of the many health keywords we have used to filter health-related tweets. 
Again, \cite{patwa2020fighting} presents a manually annotated dataset containing 10,700 social media posts and articles from various sources, such as Twitter and Facebook, and analyzes ML methods' performance to detect fake news related to COVID-19. The ML models explored in that study were Decision Tree, Logistic Regression, Gradient Boosting, and Support Vector Machine. They have not focused on health-related misinformation, which is the scope of the current study. \ignore{It is revealed that using the TF/IDF method for feature extraction, the SVM yields the best performance, achieving $93.46\%$ F1-score.} Next, \cite{gundapu2021transformer} have used supervised ML and deep learning transformer models (namely BERT, ALBERT, and XLNET) for COVID-19 misinformation detection. Likewise, in the previous one, they have not provided insights on any health-related themes or keywords. In \cite{Park2020}, an investigation on the information propagation and news sharing behaviors related to COVID-19 in Korea is performed, using content analysis on real-time Twitter data shared by top news channels.  
The results show that the spread of the COVID-19 related news articles that delivered medical information is more significant than non-medical information; hence medical information dissemination impacts the public health decision-making process.


\ignore{also compare the performance of different techniques for COVID-19 fake news detection categorized into ML, deep learning, and transformer models using a previously published dataset \cite{patwa2020fighting}. 
They found that the ensemble of three transformer models  namely BERT, ALBERT, and XLNET provides the best {accuracy}, {precision}, {recall}, and {F1-score}. 
Other studies have compares several ML algorithms with an ensemble learning to identify misinformation on Twitter related to the COVID-19\cite{Rakhami9178271}. The authors use a two-level features and study the performance of Naive  Bayes, $k$-Nearest Neighbor, Decision Tree,  Random Forest, and Support Vector Machine, where it is shown therein that the ensemble algorithm due to SVM and Random Forest produces the best detection results. }


\paragraph{\textbf{NLP for COVID-19.}}
NLP methods have also been leveraged to detect COVID-19 misinformation in YouTube ~\cite{Serrano2020,li2020youtube} and Twitter ~\cite{Rakhami9178271}. \citet{Serrano2020} have studied catching COVID-19 misinformation videos on YouTube through extracting user conversations in the comments and proposed a multi-label classifier 
Next, \citet{Rakhami9178271} use a two-level ensemble-learning-based framework using Naive  Bayes, $k$-Nearest Neighbor, Decision Tree, Random Forest, and SVM to classify misinformation based on the online credibility of the author. They define {\em credibility} based on user-level and tweet-level features leveraging NLP methods. Their findings show features like account validation ({\tt IsV}), number of retweets {\tt NoRT}), number of hashtags {\tt NoHash}), number of mentions {\tt NoMen}), and profile follow rates {\tt FlwR}) are good predictors of credibility. However, in our work, we have not used any user-level information for classifying misinformation and only relied on the texts of the corresponding tweets. These above user-level features may be added as a complement to our methodology for more accurate detection models. Again, we find study related to COVID-19 
that has leveraged machine learning algorithms with Bag of Words (BoW) NLP-based features for classifying COVID-19 diagnosis from textual clinical reports \cite{khanday2020machine}. 
 
\ignore{Focusing on videos with more than 20 comments, NLP techniques have being used to ~\cite{Serrano2020}. In this study the data was obtained by searching for YouTube videos through user-generated content on social media platforms. 
Only videos with more than twenty comments were used for this study.}

\paragraph{\textbf{Other Related Works.}} 
\citet{li2020youtube} have selected the top viewed 75 videos with keywords of `coronavirus' and `COVID-19' to be analyzed for reliability scoring. The videos have been analyzed using their proposed novel COVID-19 Specific Score (CSS), modified DISCERN (mDISCERN), and modified JAMA (mJAMA) scores. In ~\cite{Ahmed2020}, the authors highlight the drivers of misinformation and strategies to mitigate it by considering COVID-19 related conspiracy theories on Twitter where they have observed ordinary citizens as the most critical drivers of the conspiracy theories. This finding highlights the need for misinformation detection and content removal policies for social media. 


\ignore{
Some of these might be relevant and needs to be discussed in related work section. 

https://paperswithcode.com/paper/drink-bleach-or-do-what-now-covid-hera-a

https://paperswithcode.com/paper/no-rumours-please-a-multi-indic-lingual

https://paperswithcode.com/paper/artificial-intelligence-ai-in-action

https://paperswithcode.com/paper/fighting-the-covid-19-infodemic-in-social

https://journals.plos.org/plosone/article?id=10.1371/journal.pone.0247086
}


\thispagestyle{plain}
\section{Methodology and Experiment Setup}
\label{sec:methodology}

\ignore{Our study is based on the Information Operations Kill chain ~\cite{schneider2019miss}, which outlines the steps to conduct misinformation operations. The framework is based on Russian “Operation Infektion” misinformation campaign. One important characteristic of any misinformation is that it propagates using existing channels by aligning the messages to pre-existing grievances and beliefs in a group~\cite{schneider2019miss, fbi2019mis}. The messages associated with a credible source are more likely to be accepted by the audience \cite{hovland1951influence}, misinformation spread in these channels has higher prevalence and wider dissemination even after been discredited \cite{wang2019systematic} \footnote{this should go as discussion, as this are motivation of correction problem(is that correct?): what if even after the misinfo detection people don't beleive it?}. To identify groups and grievances \footnote{what groups are those? Not clear. We have not considered the source (who is the tweet owner) of the tweet in this study.}, An example of derived themes and keywords is shown in Table~\ref{themes}. {\em Themes} describe the pre-existing grievances and beliefs of the group, and {\em{ keywords}} are words specific to COVID-19 health-related misinformation. 
}

In this study, we first try to understand the types of COVID-19 health-related misinformation that have been disseminated during the pandemic.    We try to map them into the Information Operations Kill chain ~\cite{schneider2019miss} to understand the steps to conduct misinformation operations. Based on our understanding, we have studied some of the popular COVID-19 related hoaxes and misinformation articles~\cite{buzz2020misnews1, buzz2020misnews2, newsg2020trail,who2020myth} to derive various {\em{themes}} and {\em{keywords}}. {\em Themes} describe the pre-existing grievances and beliefs of the group, and {\em{ keywords}} are words specific to COVID-19 health-related misinformation that align with a particular {\em theme}. These keywords help us to collect and filter relevant tweets from a sheer volume of COVID-19 related tweets for the current study. Again, we have selected only a few days to collect Twitter data for resource and time constraints. However, to increase the chances of coverage of common COVID-19 health-related tweets, we have selected dates where a COVID-19 related event has occurred in the United States. This pilot study would reveal the efficacy of our proposed methodology. The selection of dates also covered the tweets from the first two months (March-April 2020) to the first four months (June-July 2020) of the global pandemic declaration on March 11, 2020 \cite{Cucinotta_Vanelli_2020}. In this section, we present our methodology for the following modules: (i) Twitter Dataset Collection, (ii) Annotation of Tweets, (iii) Analyzing Tweets, (iv) Classification Tasks for Detection Model, (v) Performance Evaluation. 



\ignore{Usually, these events were commonly targeted for misinformation spread to divide people and bring chaos in society.}

\begin{table*}
\resizebox{\textwidth}{!}{
 \begin{tabular}{ P{3.25cm} P{3.25cm} P{3.25cm} P{3.25cm} P{3.25cm} P{4cm}}
 \toprule
 \multicolumn{1}{c}{\textbf{Theme: Limiting Civil Liberties}} & \multicolumn{3}{c}{\textbf{Theme: Prevention}} & \multicolumn{2}{c}{\textbf{Theme: Possible Remedies}} \\
 \cmidrule(lr){1-1} \cmidrule(lr){2-4} \cmidrule(lr){5-6}
 lockdown & tide pods	 & hand sanitizer & immunity drink & hydroxychloroquine & homeopathy \\
 hoax & sunlight & uv ray & uv light & chloronique & bleach shots\\
  &	  & desinfectant &  &  & \\ \midrule \midrule
  & \multicolumn{2}{c}{\textbf{Theme: Worsening Condition}} & \multicolumn{2}{c}{\textbf{Theme: Origin of the Virus}} & \\
   \cmidrule(lr){2-3} \cmidrule(lr){4-5} 
  & amphetamine & 5g & wuhan virus & wuhan virus & \\
 \bottomrule
\end{tabular}}
\caption{Example of a COVID-19 Health Related Themes and Keywords}
\label{tab:themes}
\end{table*}

\begin{table*}
\resizebox{\textwidth}{!}{
  \begin{tabular}{ P{3cm}  P{3cm} P{3cm} P{3cm} P{3cm} P{3cm} P{3cm}}
 \toprule
 \multicolumn{7}{c}{\textbf{COVID-19 Health Related Keywords}} \\
 \cmidrule(lr){1-7} 
 drug &	dengue & wash hand & antibody & fda-approved & facemask & sars\\
 treatment	& screening patient & hand wash & immunity & scientific evidence & health & vaccine\\
 stay at home &	home stay  & social distance & azithromycin & mask & stay home & fever\\
 testing & immunity drink & heperan sulfate & pneumonia & vitamin d & body pain & n95\\
 slight cough & n-95 & herd immunity & antibody test & antibodies &  &  \\
 \bottomrule
\end{tabular}}
\caption{List of COVID-19 Health-Related Keywords for Filtering Tweets}  
\label{info_kw}
\end{table*}

\subsection{Twitter Dataset Collection}
\label{sec:dataset}
We only consider collecting the tweets or user-generated content posted by anonymous individuals from {\em Twitter}. This study does not infer or reveal any authors of the tweets, as we only extract the textual parts of a tweet. The biggest challenge for collecting quality tweets is similar to a ``needle in haystack'' problem as there are many COVID-19 related tweets on Twitter posted daily. 
We have focused only on health-related tweets because that directly impacts public health if people trust misguided tweets. 


Although it is possible to collect the related tweets directly using Twitter API via Tweepy~\cite{roesslein2020tweepy}, there are rate limitations on the API. 
As an alternative in this study, we use the COVID-19 Tweets dataset~\cite{lamsal_2021} from IEEE Dataport because this dataset is (i)  publicly available online, (ii) provides COVID-19 related tweet IDs daily from as early as March 2020, and (iii) the tweets are collected from all over the world (thereby giving no regional limitation). 

We have selected the following four days: 04/02/2020 (i.e., Stay-at-Home orders in many states in the US), 04/24/2020 (i.e., POTUS comments on the use of disinfectant against COVID-19 goes viral), 06/16/2020 (i.e., reports published on the use of common drugs against COVID-19~\cite{commondrugs_covid_2020}), and 07/02/2020 (i.e., face cover mandates in many US states). After selection of the dates, we download the full dataset for each of the dates from IEEE Dataport \cite{lamsal_2021}. The data contains all the tweet IDs for the tweets, but it does not contain the tweets (i.e., texts) themselves. Next, we use the Hydrator app \cite{documentingthenow} for extracting 
actual tweets. For each selected day, we extract 10,000 tweet IDs and collect those tweets for further processing. We limit our collection to 10,000 tweets because of resources and time limitations. We observe that the tweets extracted from the Hydrator app are truncated to a maximum of 280 characters. 

Next, we identify various themed for ongoing COVID-19 health-related misinformation (as shown in Table \ref{tab:themes}) and define a glossary of COVID-19 health-related keywords (shown in Table 
\ref{info_kw}) to filter only the interesting and relevant health-related tweets. This filtering resulted in a total of 2,615 unique tweets for all four selected dates. The health keyword glossary combines COVID-19 health-related misinformation (based on the themes) and true information. 

\thispagestyle{plain}
\subsection{Annotations of Tweets} 
\label{sec:data_labeling}
\ignore{Data is collected from tweet IDs available on IEEE dataport for the following four days: 04/02/2020, 04/24/2020, 06/16/2020, and 07/02/2020. \textbf{First,} The Hydrator app \cite{documentingthenow} is used to extract 10,000 tweets for each of the selected days. We limit our collection to 10,000 tweets because of resources and time limitations. We find that the collected tweets are truncated to a maximum of 280 characters. \textbf{Second,} a glossary of 50 COVID-19 health-related words (Tables \ref{themes} \& \ref{info_kw}) is built to retains only the interesting and relevant tweets, which resulted in 2,615 unique tweets (i.e., unique strings). The glossary was a combination COVID health related misinformation (based on the themes and keywords mentioned above) and information words.}  

In this study, we apply manual annotation on the filtered tweets to label them. 
Initially, we have defined 5 class labels to annotate all the filtered tweets, such as  (i) true information ($T$), (ii) misinformation ($M$), (iii) incomplete ($I$), (iv) not health-related ($N$), and (v) unsure ($U$).
Here, {\em true information} is COVID-19 related health facts supported by scientific evidence; {\em misinformation} is inaccurate COVID-19 health-related information that health organizations like WHO and CDC have discredited; {\em incomplete information} is a truncated tweets that can not be verified for the complete statement; {\em not health-related} is any tweet about COVID-19 that does not directly relate to any health information; and {\em unsure} class contains tweets where the annotator is unsure about the exact categorization to any of the first four labels. 
\ignore{First, each of the team members independently read through the tweets and label them based on the commonly agreed terms. We have selected four labels to annotate{\footnote{\color{magenta}From Dr Rios: We need to describe the annotation agreement}} the tweets: Next, we conduct a majority voting to select the final annotated labels for each tweet. }

The same set of tweets are independently labeled by multiple annotators (i.e., our research team members). We have relied on majority voting among the tweet labels to finalize labels for each tweet. Moreover, any of the tweets not having a winning label are finalized by conducting open discussion among the group of annotators to reach a unanimous agreement. This process has produced 314 tweets labeled as $T$, 210 tweets labeled as $M$, 173 tweets labeled as $I$, and 1,918 tweets labeled as $N$. \ignore{The flowchart depicting all the steps involved in data set preparation for training and testing is shown in Figure \ref{fig:Data}.} For aiding the future research on COVID-19 health-related misinformation detection on social media, we would be happy to share our annotated ground-truth dataset through valid organizational email requests only. 
We believe the dataset can work as a basis for improved future research. 

\subsection{Methods for Analyzing Tweets}
\label{sec:method_for_tweet_analysis}
We have only considered using the tweets with class labels $M$ and $T$ for health-related misinformation detection study. At first, we tokenize  tweets to build two sets of tokens: (i) {\em true information} tokens, (ii) {\em misinformation} tokens. Next, to improve the model performance, we remove the default English stop words (${\sf SW}_{english}$) listed in ~\cite{english_stopwords} and more trivial COVID-19 words for true information and misinformation tweets. These trivial words are observed by analyzing the most frequent tokens from {\em true information} and {\em misinformation} set of tweets. Some of the highlighted trivial {\em COVID-19 words} are included in this set ${\sf SW}_{trivial}$ = \{{\tt covid19}, {\tt covid}, {\tt covid-19}, {\tt coronavirus}, {\tt corona}, {\tt covid\_19}, {\tt health}\}. Moreover, it is important to cleanup tweets for reliable and effective analysis as many tweets are messy. The cleanup process includes transmitting all tweets to lower case letters for avoiding redundancy, removal of incomplete links, cleaning of unnecessary punctuation, irrelevant set of characters (e.g., ..., “,”, @), non-meaningful single-character tokens, and digits only tokens (e.g., ``100''). In total, we have removed 222 stop-words (presented as {\sf SW}, where ${\sf SW} = {\sf SW}_{english} \cup {\sf SW}_{trivial}$). Next, we apply python NLTK SnowballStemmer for stemming the tweets \cite{nltk_doc} and extract the root forms of words for more generalized model. 
After these steps, we need to generate features from the tweet texts for the supervised learning classifiers. This study only relied on text features (e.g., extracted from individual tweets) to classify health misinformation versus true information. In this study, we have used popular Bag of Words (BoW) ~\cite{Zhang2010} and different n-grams~\cite{Furnkranz98astudy} NLP methods for feature extraction from the tweets. 

\thispagestyle{plain}
\subsubsection{Feature Extraction}
\paragraph{\bf Bag of Words (BoW)}: This method uses raw word frequencies in a sentence as features. If the {\em BoW} method is used, a tweet is presented as set of tokens, where each token is a connected word (e.g., no spaces in between), and it stores a frequency count for that token in the tweet. 
Any tweet $tw_i$ containing multiple BoW tokens (or word) where each token, ${\sf tok}_j = w_j \notin {\sf SW}$, and class label of the $i$-th tweet, $l_i\in \{M,T\}$ can be presented as a set of BoW tokenized representation ${\sf tokenized}_i^{B} = {\sf P}_{i}^{B} \cup {\sf A}_{i}^{B}$ for the $i$-th tweet. Now, ${\sf P}_{i}^{B}$ is the set of tokens that are present and ${\sf A}_{i}^{B}$ is the set of tokens absent in the $i$-th tweet $tw_i$, derived as follows  
\ignore{
\begin{equation}
\label{eq:bow_tokenize}
    {\sf tokenized}_{i} = \{ \forall_{j}{\sf tok}_j = {\sf Freq({\sf tok}_j)} : {w}_j \in tw_i\}
\end{equation}
}
\begin{equation}
\label{eq:bow_tokenize_present}
    {\sf P}_{i}^{B} =
    \{\forall_{j}{\sf tok}_j := {\sf Freq({\sf tok}_j)} \text{ | } {w}_j\in{tw_i}\}
\end{equation}

\begin{equation}
\label{eq:bow_tokenize_absent}
    {\sf A}_{i}^{B} =
    \{\forall_{j}{\sf tok}_j := 0 \text{ | } {w}_j \notin{tw_i}\}
\end{equation}
In Eq. \ref{eq:bow_tokenize_present} and \ref{eq:bow_tokenize_absent}, ${\sf Freq}({\sf tok}_j)$ present the frequency counts of any $j$-th token within the $i$-th tweet.

\paragraph{\bf n-grams}: This method uses the sequence of $n$ (where $n$=\{1, 2, 3\}) words as binary features. 
For \textbf{$1$-gram} (or {\em \textbf{uni-gram}}) a single word sequence is considered as a token. Unlike BoW method, $1$-gram method uses features for each token as binary values (1 or 0), stating whether the token is present in a tweet or not and does not mention if there are multiple instances of the token. 
Next, we have used \textbf{$2$-grams} (or {\em \textbf{bi-grams}}) as features, which use all the sequences of two words as tokens and stores them with 1 or 0 binary values. Lastly, the \textbf{$3$-grams} (or {\em \textbf{tri-grams}}) features use all the valid sequences of three words as tokens with the binary value 1 (if token is present in tweet) or 0 (if token is not present in tweet).

Now, any tweet $tw_i$ containing all $n$-grams tokens $\forall_{j}{\sf tok}_j^{n}$, where, $n\in \{1,2,3\}$ and class label of the $i$-th tweet, $l_i\in \{M,T\}$ can be presented as a set of $n$-grams tokenized representation, ${\sf tokenized}_{i}^{n} = {\sf P}_{i}^{n} \cup {\sf A}_{i}^{n}$ for the $i$-th tweet. Here, ${\sf P}_{i}^{n}$ and ${\sf A}_{i}^{n}$ present the set of tokens that are present and absent in tweet $tw_i$, which is derived by Eq. \ref{eq:ngram_tokenize_present} and Eq. \ref{eq:ngram_tokenize_absent}, respectively. 
\begin{equation}
\label{eq:ngram_tokenize_present}
    {\sf P}_{i}^{n} = 
    \{\forall_{j}{\sf tok}_j^n := 1 \text{ | } \forall_{w \in {\sf tok}_j^n} w \in{tw_i}\}
\end{equation}
\begin{equation}
\label{eq:ngram_tokenize_absent}
    {\sf A}_{i}^{n} =
    \{\forall_{j}{\sf tok}_j^n := 0\text{ | } \forall_{w \in {\sf tok}_j^n} w \notin{tw_i}\}
\end{equation}
In Eq. \ref{eq:ngram_tokenize_present}, $\forall_{w\in {\sf tok}_j^n} w \in tw_i$ depicts the presence of all the words of the $j$-th token in tweet $tw_i$. Now, $n$-gram tokens is further derived based on the value of $n$, as follows, 

\begin{equation}
\label{eq:ngram_tokens}
    {\sf tok}_{j}^{n} =
\begin{cases}
    (w_j) & \text{if } n =1 \\
    (w_j, w_{j+1}) & \text{if } n =2 
    \\
    (w_j, w_{j+1}, w_{j+2}) & \text{if } n = 3
\end{cases}
\end{equation}

From equation \ref{eq:ngram_tokens}, we see that {\em uni-gram} method ($n$=1) use single word $w_j\in tw_i$ as tokens for any $tw_i$. The {\em bi-grams} method ($n$=2) use a pair of two words $(w_j, w_{j+1})$ as tokes where both $w_j, w_{j+1}\in tw_i$. Lastly, {\em tri-grams} method ($n$=3) use a tuple of three words $(w_j, w_{j+1},w_{j+2})$ as tokes, where $w_j, w_{j+1}, w_{j+1}\in tw_i$. Moreover, for any tweets containing $J$ number of words has ($J - n + 1$) tokens for any $n$-grams methods.

\ignore{
Now, any tweet $tw_i$ containing $K$ words $\forall_{j\in [1,K]} w_j\notin {\sf SW}$, and label $l_i\in \{M,T\}$ can be presented by the following equation of tokens ${\sf tok}_i$ for the $i$-th tweet, 
\begin{equation}
    {\sf tok}_{i} = \{ \forall_{j}w_j = True: w_j \in tw_i\}
\end{equation}
}
\subsubsection{Preparing Training and Test Data}
The selection of the feature extraction method plays a critical role in health-related misinformation detection. 
To start preparing the training and testing datasets, the set of tokenized {\em misinformation} tweets is presented as $D_{M} = \{\forall_{i} {\sf tokenized}_{i}^{m\in\{B,n\}} | \text{ } l_i=M\}$, while the the set of tokenized {\em true information} tweets are presented as $D_{T} = \{\forall_{i} {\sf tokenized}_{i}^{m\in\{B,n\}} | \text{ } l_i=T\}$. Here, $m\in\{B,n\}$ representing the method for feature extraction. Next, we merge these dataset to prepare as $D_{merge} = D_{M} \cup D_{T}$. Then, the $D_{merge}$ dataset is randomly splitted with $80:20$ ratio into training data ($D_{train}$) and test data ($D_{test}$), along with their respective tweet labels. 
In this pilot study, our training dataset contains $|D_{train}|=419$ tweets, of which 164 are {\em misinformation} and 255 are {\em true information}. Again, the testing dataset contains $|D_{test}|=105$ tweets, of which 46 are {\em misinformation} and 59 {\em true information}. Note that, individual tweets shuffle from training to test and vise versa, but the training and test size remain constant, which causes a change in the number of tokens in training and test between BoW and 1-gram method. 

\subsubsection{Analysis with BoW Features}
For the {\em BoW} method, we observed $5,268$ words (or {\em tokens}) in $D_{train}$ (training data), leading to a vocabulary size of $2,301$ unique words, and $1,318$ words in $D_{test}$ (test data), leading to a vocabulary size of $838$ unique words. Moreover, the tweets contained in $D_{train}$ are consisting 12.57 words on average, with maximum of 32 and minimum of 2 words. Likewise, tweets contained in $D_{test}$ are also consisting 12.55 words on average, with maximum of 28 and minimum of 2 words, which indicates a similar distribution of tweets in both training and testing datasets with this method.

\subsubsection{Analysis with $n$-grams Features}
In case of {\em uni-gram} method, we have 5,232 tokens in $D_{train}$ leading to 2,286 unique {\em uni-gram} tokens, and 1,354 tokens in $D_{test}$ leading to 848 unique {\em uni-gram} tokens. The average token length for training set is 12.49 tokens, with maximum of 32 and minimum of 2 tokens, while the average token length for test set is 12.90 tokens, with maximum of 30 and minimum of 4 tokens. Next, in case of {\em bi-grams}, we have 4,813 tokens in $D_{train}$ leading to 4,298 unique {\em bi-grams} tokens, and 1,249 tokens in $D_{test}$ leading to 1,171 unique {\em bi-grams} tokens. The average token lengths for training set is 11.49 tokens, with maximum of 31 and minimum of 1 tokens, while the average token length for test set is 11.90 tokens, with maximum of 29 and minimum of 3 tokens. Lastly, in case of {\em tri-grams}, we have 4,394 tokens in $D_{train}$ leading to a {\em tri-grams} vocabulary (unique tokens) size of 4,166, and 1,144 tokens in $D_{test}$ leading to vocabulary size of 1,109. The average token length for training set is 10.49 tokens, while the average token lengths for test set is 10.90 tokens.


\begin{figure*}[!htpb]
\centering
\includegraphics[width=1\textwidth]{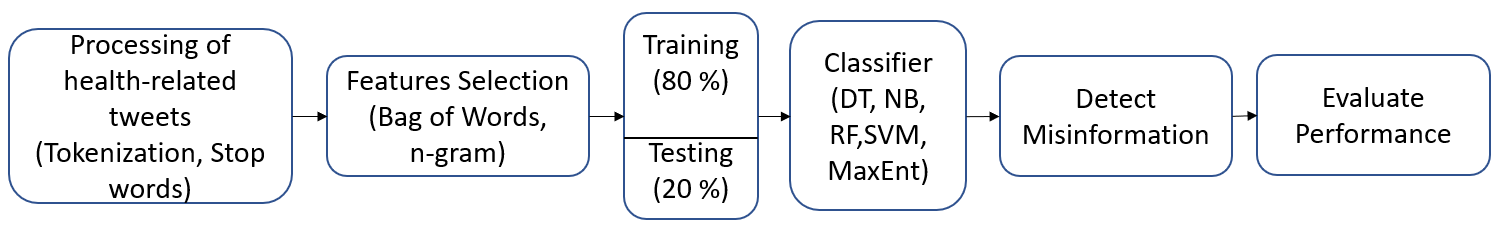}
\caption{Steps for Processing of Health-Related Tweets to Detect {\em Misinformation}}
\label{classifier}
\end{figure*} 
\thispagestyle{plain}
\subsection{Classification Tasks for Misinformation Detection} 
The purpose of the classification task is to separate COVID-19 health-related tweets between {\em true information} and {\em misinformation}. We have used multiple machine learning classifiers that are commonly used in literature for text classification. The present study considers Decision Tree (DT)~\cite{Song2015}, Naive Bayes (NB)~\cite{LangleyNaiveBayes}, Random Forest (RF)~\cite{biau12a}, Support Vector Machine (SVM)~\cite{Evgeniou2001}, and Maximum Entropy Modeling (MEM)~\cite{MaxEntropyModeling_Berger} classifiers. We have selected these classifiers because, (i) they are popular, (ii) they are readily available under the python {\tt nltk.classify} and {\tt nltk.classify.scikitlearn} API libraries ~\cite{nltk_doc}, (iii) they would provide a basis if the proposed methodology is a viable way of tackling the COVID-19 themed health-related misinformation. These classifiers are trained based on the input features extracted from the tweets (e.g., tokens) and corresponding numerical labels assigned using the manual annotation process described in section \ref{sec:data_labeling}. The steps involved in training the classification models are shown in Fig. \ref{classifier}. 

\subsubsection{Performance Metrics}
We have used the following standard evaluation metrics to evaluate our detection (i.e., classification) methodology: {\sf class-wise Precision}, {\sf class-wise Recall}, {\sf class-wise F1-score}, {\sf classification Accuracy}, and {\sf Macro-F1-score}. 
The following list provides the formulas for each of the metrics:


\begin{itemize}
    \item $    \text{{\sf Accuracy}} = \frac{1}{2}\sum_{c \in \{M, T\}} {\sf Accuracy}_{c}$, where individual class accuracy ${\sf Accuracy}_{c} =   \frac{\text{TP}_c+\text{TN}_c}{\text{TP}_c+\text{TN}_c+\text{FP}_c+\text{FN}_c}$.
    \item $\text{\sf Precision}_{c} = \frac{\text{TP}_c}{\text{TP}_c+\text{FP}_c}$, for the selected class $c \in \{M,T\}$.
    \item $    \text{\sf Recall}_{c} = \frac{\text{TP}_c}{\text{TP}_c+\text{FN}_c}$, where $c\in \{M,T\}$ depicts the class. 
    \item $\text{\sf F1-score}_{c\in \{M,T\}} = \frac{2\times({\sf Recall}_c \times {\sf Precision}_c)}{{\sf Recall}_c + {\sf Precision}_c}$.
    \item $\text{\sf Macro-F1-score} = \frac{1}{2}\sum_{c \in \{M, T\}} \text{{\sf F1-score}}_c$.

\end{itemize}
Here, ${\text{TP}}_c$ is the total number of true positives, ${\text {TN}}_c$ is the total number of true negatives, ${\text{FP}}_c$ is the number of false positives, and ${\text{FN}}_c$ is the total number of false negatives for class label $c\in \{M,T\}$. 

\begin{table*}[ht]
\centering
\renewcommand{\arraystretch}{1.05}
\resizebox{\textwidth}{!}{%
\begin{tabular}{llrrrrrrrr}
\toprule
& & \multicolumn{3}{c}{Misinformation} & \multicolumn{3}{c}{True information} & \multicolumn{2}{c}{Aggregated Metrics} \\ \cmidrule(lr){3-5} \cmidrule(lr){6-8} \cmidrule(lr){9-10}
                     &  & Precision & Recall & F1-score    & Precision & Recall & F1-score    & Accuracy & Macro-F1 \\ \midrule
\multirow{4}{*}{NB} & {\em BoW} & .684     & .650  & .667 & \textbf{.785}     & .810  & .797 & 0.65    & \textbf{.732}\\
 & {\em uni-gram} & .683     & \textbf{.683}  & \textbf{.683} & .780     & .780  & .780 & \textbf{0.74} & .731 \\
 & {\em bi-grams}   & .680     & .472  & .557 & .747     & .875  & \textbf{.806} & 0.73 & .682 \\
 & {\em tri-grams}  &  \textbf{1.0}    & .211  & .348 & .674     & \textbf{1.0}  & .805 & 0.70 & .577 \\ \midrule
\multirow{4}{*}{DT} & {\em BoW}                     & .600     & \textbf{.585}  & .593 & .730     & .742  & .736 & 0.67 & .664\\
 & {\em uni-gram}                    & .852     &  .561 & \textbf{.677} &  \textbf{.753}    & .932 & \textbf{.833} & \textbf{0.78} & \textbf{.755} \\
 & {\em bi-grams}                     & \textbf{1.0}     & .278  & .435 & .711     & \textbf{1.0}  & .831 & 0.74 & .633 \\
 & {\em tri-grams}                     & 1.0     & .131  & .233 & .653     & 1.0  & .790 & 0.67  & .511 \\ \midrule
\multirow{4}{*}{MEM} & {\em BoW}                    & .636     & .512  & .568 & .714     & \textbf{.807}  & .758 & 0.68 & .663\\
 & {\em uni-gram}                    & \textbf{.692}     & .659  & \textbf{.675} & .771     & .797  & \textbf{.783} & \textbf{0.74} & \textbf{.731} \\
 & {\em bi-grams}                    & .492     & .861  & .626 & \textbf{.865}     & .500  & .634 & 0.63 & .630 \\
 & {\em tri-grams}                    & .413     & \textbf{.868}  & .559 & .750     & .242  & .366 & 0.48 & .463 \\ \midrule
\multirow{4}{*}{RF} & {\em BoW}                     & .700     & .342  & .459 & .675     & .903  & .772 & 0.67 & .616 \\
 & {\em uni-gram}                     & \textbf{.895}     & \textbf{.415}  & \textbf{.567} &   \textbf{.704}   & .966  & \textbf{.814} & \textbf{0.74} & \textbf{.691} \\
 & {\em bi-grams}                     & .889     & .222  & .356 & .692     & .984  & .813 & 0.71 & .533 \\
 & {\em tri-grams}                     & .653     & .132  & .233 & .653     & \textbf{1.0}  & .790 & 0.67 & .511 \\ \midrule
\multirow{4}{*}{SVM} & {\em BoW}                    & .737     & .341  & .467 & .679     & .919  & .781 & 0.69 & .624 \\
 & {\em uni-gram}                    & .739     & \textbf{.415}  & \textbf{.531} & .688     & .898  & .779 & 0.70 & .655 \\
 & {\em bi-grams}                    & .923     & .333 & .490 & \textbf{.724} & .984  &  \textbf{.834} & \textbf{0.75}& \textbf{.662} \\
 & {\em tri-grams}                    &  \textbf{1.0}  & .105  & .191 & .646     & \textbf{1.0}  & .785 & 0.66 & .488 \\ \bottomrule
\end{tabular}%
}
\caption{Performance of Baseline Classifiers for Misinformation, True Information classes, and Aggregated Metrics (test data size = 105)} 
\label{tab:result-table}
\end{table*}

\section{Experiment Results and Discussions} 
\label{sec:discuss}




Table \ref{tab:result-table} shows that the {\em uni-gram} text representation method with the Decision Tree classifier achieves the best classification accuracy of 78\%. We also observe that {\em uni-gram} method has outperformed all other methods for all the classifiers except SVM, where the {\em bi-gram} method has 75\% over the 70\% accuracy of the {\em uni-gram} method. Table \ref{tab:result-table} also highlights that all the classifiers achieve at least 74\% classification accuracy with at least one of the text representation methods. We also looked at the class-wise F-1 score for both misinformation class and true information class, which further validates that uni-gram method is outperforming other n-grams and BoW methods consistently for misinformation class labels. The classifiers perform much better for detecting {\em true information} labeled tweets with F1-score ranging from 0.783 to 0.833 (i.e., at least one of the text representation methods) for different classifiers. In contrast, F1-score for misinformation detection is always higher for the uni-gram method, with a maximum of 0.683 for the NB classifier. One reason for getting a better result for the true information than the misinformation class is the imbalance ratio 1:0.686 for {\em true information}:{\em misinformation} in the dataset (see data distribution for both classes in section \ref{sec:method_for_tweet_analysis}). Moreover, in reality, the number of tweets with true information is usually much higher than the actual misinformation in social media, which is imitated in this study. We also observe that the highest Macro-F1-Score of 0.755 is achieved for the Decision Tree classifier with the {\em uni-gram} method, which further indicates the effectiveness of {\em uni-gram} modeling for health-related tweets' classification. 

Next, for Precision and Recall, we find that some of the methods (e.g., {\em bi-grams}, {\em tri-grams}) have achieved a perfect precision (1.0) or recall (1.0) for either of the class. However, whenever a classifier achieved a perfect precision (very few false positives) for the misinformation class, it has a significantly lower recall of around 0.2 (very high false negatives). Any such behavior would count as a bias classifier towards one of the class labels. Hence, we want to choose prevision and recall more balanced for both of the class labels. Another insight we draw from the current analysis is that the {\em bi-grams} and {\em tri-grams} methods have not been performing well enough because most of the {\em bi-grams} and {\em tri-grams} tokens are unique and are not repeated in the dataset for both class labels. We also feel that the small sample data size $(n=524)$ including both the training and testing phase, is also not enough for a rigorous study but it is certainly the first step towards exploring the significance of the problem. We believe with a larger dataset, our methodology's performance could be further generalized.

\thispagestyle{plain}
\subsection{Limitations}
There are still some limitations in the study. \textbf{First}, we are not inferring the tweet's meaning. 
For example, a classifier's failure would look like classifying two-sentences, {\tt sentence-1}=``Hydroxychloroquine is a medicine for COVID-19'', and {\tt sentence-2}=``Hydroxychloroquine is not a medicine for COVID-19''. Though {\tt sentence-2} $= \neg ${\tt sentence-1}, and {\tt sentence-2} is {\em true information}, it does not have enough samples in the training dataset, and may get classified as {\em misinformation}. Thus, to handle such scenarios, we need to build further methods to get negation versus affirmation meaning from a tweet to get more accurate models. We can also include anti-misinformation (i.e., true information countering the existing misinformation) in our training datasets to make the model more proactive in detecting misinformation campaigns. 
\textbf{Second}, we have only selected four dates and 10,000 tweets for each day due to the lack of time and resources to invest in the manual annotation process. 
Moreover, some of the tweet IDs (around 5-10\% of 10,000 tweets each day) we have extracted empty tweets (i.e., removed by Twitter or author). 
\textbf{Third}, the lack of ground-truth datasets forced us to do manual labeling of the data to train supervised learning-based classifiers. These manual tasks may have impurity and errors, but we have used our best judgments and followed the best practices to apply various mechanisms (e.g., group discussion, majority voting) to bring fairness into the annotation process. \textbf{Fourth}, We have only considered BoW and $n$-gram methods for this pilot study and have not examined other methods such as TF-IDF and word embedding. Moreover, using the BoW method has its own shortcomings \cite{bow_limitation} which is also present in the current study. \textbf{Fifth}, misinformation through images and videos are not focused in this study. \textbf{Sixth,} we only analyze Twitter as social media platform, but we believe the methodology is still applicable to other platforms (e.g., Facebook, Instagram) with minimal tweaks. However, multimedia-based platforms (e.g., TikTok) need different approaches. 

\thispagestyle{plain}
\subsection{Ethical Considerations} 
Although 
it is essential to understand how to mitigate effects of misinformation, there are some ethical considerations. Misinformation is commonly encountered in the form of selective information or partial truths in conversations about controversial topics \cite{schneider2019miss}. An example is the statistics on {\em Black-on-Black crimes} that are used to explain over policing of Black communities \cite{braga2015police}.  
From this perspective, misinformation can be generated when the information available is incomplete (e.g.in news developing stories), or when a new findings appear contradicting existing beliefs~\cite{morawska2020airborne}. 
We think misinformation detection mechanisms must consider these factors to avoid being viewed as censorship or violation of freedom of speech~\cite{kaiser2020adapting}. To ensure freedom of speech, we have opted to label tweets as {\em misinformation} if we feel the author of any tweet is demanding questions on the existing health system, therapeutics, or policies for tackling this pandemic. We have also ensured not to violate the privacy of any Twitter users by not using any sensitive account information about any tweet's owner (author of any tweet labeled as $T$ or $M$). We have also cleaned up our tweets using $regex$ where texts contain tagging other Twitter users with `@TwitterID' tags. Lastly, we believe that a better policy for addressing misinformation would be to provide correct information that does not threatens individuals' existing beliefs \cite{chan2017debunking} but deter them from harmful behavior instead blocking misinformation contents. We recommend that future studies should investigate the solution space in this direction for systematizing it. 

\subsection{Future Directions}
In future, it would be interesting to explore detection mechanisms with explainable classifiers for bringing trustworthiness to misinformation detection research. Another future work arena would be to study network activities such as retweets, likes, and followers count for health misinformation tweet accounts to see if bots or real accounts have disseminated it. An analysis of those network elements could validate existing misinformation propagation frameworks \cite{fbi2019mis} for health-related misinformation, which could be leveraged to develop proactive mechanisms to detect unseen misinformation activity in real-time. Finally, along with the detection part, correcting health-related misinformation among communities need to be studied. Because of the continued influence effect, misinformation once presented continues to influence later judgments, the author proposed compelling facts as a medium of correction ~\cite{correction_cont_influence}, which needs to be further analyzed in the context of COVID-19 health misinformation.

\section{Conclusion}
\label{sec:conclusion}
In this paper, we present a methodology for health misinformation detection in Twitter leveraging state-of-the-art techniques from NLP, ML, misinformation propagation, and existing social psychology research domains. We find extracting and annotating quality data for misinformation research is challenging. We discover gap in availability of ground-truth dataset, which is addressed by our effort. Our findings highlight Decision Tree classifier is outperforming all other classifiers with 78\% {\em classification accuracy} leveraging simpler {\em uni-gram} features for text representation. We recommend future studies to systematize the understanding of the health-related {\em misinformation} dissemination in social media and its impact on public health at large.
\thispagestyle{plain}

\ignore{
Bag of Words:Training set is small to reflect actual frequency distribution of tokens,  do not have knowledge about the negation or assertion in sentence. Only based on the token frequency.

1-gram:Training set small for containing good distribution of words for both true and misinformation, may depends on the stop words, which need to be analysed.
2-gram:Training set small for containing good distribution of words for both true and misinformation, may depends on the stop words.
3-gram:Training set small for containing good distribution of words for both true and misinformation, may be 3 token's sequence is longer considering the average tokens in tweets. that's why unknown tweets in the test set couldn't be identified correctly. 
conclusions goes here ....

Esra'a 

Determining engineering features:  point 3 in the literature review list: Splitting all correlative COVID-19 posts of into several groups to extract all the features by modeling these groups. This is a good reference for detecting misinformation in general: Yu, F., Liu, Q., Wu, S., Wang, L., \& Tan, T. (2017). A convolutional approach for misinformation identification.

}
\thispagestyle{plain}

\bibliography{anthology,bib_file}
\bibliographystyle{acl_natbib}

\thispagestyle{plain}
\end{document}